\documentclass[twocolumn,showpacs,preprintnumbers,amsmath,amssymb,prl,epsf,superscriptaddress]{revtex4}
\usepackage{graphicx}
\usepackage{nicefrac}
\usepackage[squaren]{SIunits}

\begin{document}
\title{Two-Photon Polarization Interference for Pulsed SPDC in a PPKTP Waveguide}
\author{Malte Avenhaus}
\affiliation{Max-Planck Research Group, G\"unther-Scharowsky-Stra\ss{}e 1/Bau 24, 91058 Erlangen, Germany}
\author{Maria~V.~Chekhova}
\affiliation{Max-Planck Research Group, G\"unther-Scharowsky-Stra\ss{}e 1/Bau 24, 91058 Erlangen, Germany}
\affiliation{Department of Physics, M.V.Lomonosov Moscow State University, \\ Leninskie Gory, 119992 Moscow, Russia}
\author{Leonid~A.~Krivitsky}
\affiliation{Max-Planck Research Group, G\"unther-Scharowsky-Stra\ss{}e 1/Bau 24, 91058 Erlangen, Germany}
\affiliation{Technical University of Denmark,
Fisikvej 309, 2800 Lyngby, Denmark}
\author{Gerd Leuchs}
\affiliation{Max-Planck Research Group, G\"unther-Scharowsky-Stra\ss{}e 1/Bau 24, 91058 Erlangen, Germany}
\author{Christine Silberhorn}
\affiliation{Max-Planck Research Group, G\"unther-Scharowsky-Stra\ss{}e 1/Bau 24, 91058 Erlangen, Germany}
\begin{abstract}

We study the spectral properties of Spontaneous Parametric Down Conversion
in a periodically poled waveguided structure of KTP crystal
pumped by ultra-short pulses. Our theoretical analysis
reveals a strongly multimode and asymmetric structure of the
two-photon spectral amplitude for type-II SPDC. Experimental evidence, based on Hong-Ou-Mandel polarization interference with narrowband filtering,
confirms this result.
\end{abstract}
 \maketitle \narrowtext
\vspace{-10mm}

During the last two decades generation and application of entangled states
of light became a critical issue in both fundamental research and practical
applications~\cite{NC}, including the implementation of the quantum
information protocols ~\cite{Kok}, realization of quantum cryptography systems
~\cite{gis}, quantum metrology ~\cite{alan} and many others. These actively
developing areas of modern physics motivate a continuous search for highly
efficient sources of entangled states.

One of the most accessible ways to generate
entangled photon states is by means of Spontaneous Parametric Down
Conversion (SPDC)~\cite{DNK}.
In the last two decades several efficient
sources of entangled states based on SPDC in bulk crystals have been proposed.
Among them are schemes based on type-II SPDC in a
single crystal with specific arrangements of the spatial modes
~\cite{sergienko,Kwiat_rings,kurt} and sources composed of two
type-I crystals with orthogonal orientation of the optic axes
~\cite{Kwiat,kulik}.

Further significant improvement of SPDC-based
sources is related to the implementation of waveguided periodically poled (PP)
structures in nonlinear crystals. In a specific production process, periodical
variation is applied to the components of the quadratic susceptibility tensor and
the interaction volume of the pump with the nonlinear crystal is restricted to a
small waveguided region fabricated on the surface of the crystal
~\cite{gisinPPLN, Walmsley, fiorentino, uren}.
In this case the phasematching relations are derived taking into account
the periodicity of the quadratic susceptibility and the effective
change of the refractive index due to the waveguided propagation
~\cite{deltaNinWG}. Highly efficient PDC in such kind of structures is
possible due to the ability to apply any desirable poling period in order to get the benefit
of the optimal
phase-matching conditions and at the same time to involve the most efficient
component of the quadratic susceptibility tensor. Moreover, due to the
waveguided propagation of the pump beam in the nonlinear crystal, the power density reaches extremely high
values (up to tens \giga\watt\per\centi\meter\squared) all over the length of the
nonlinear waveguide. So far, PDC generation in periodically poled nonlinear
waveguides has been demonstrated both for type-I operation (in $\text{PPLiNbO}_3$)
~\cite{gisinPPLN} and for type-II operation (in PPKTP)
~\cite{Walmsley, fiorentino}. In both cases a significant improvement of PDC
efficiency was achieved in comparison with the traditional sources.

At the same time, several properties of SPDC in periodically poled waveguides remain unexplored. In particular, these are the spectral properties of the state produced at the output. Due to the large length of the waveguide in combination with the short pulse duration, the state is highly entangled in frequency~\cite{Fedorov} and the two-photon spectral amplitude (TPSA) has an extremely narrow width of conditional (coincidence) spectral distribution. The straightforward method of measuring TPSA by simultaneously filtering and scanning the frequencies of signal and idler photons~\cite{Antia} is not feasible here because it requires optical filters with FWHM much less than the TPSA width. In this paper, we study the TPSA of type-II SPDC generated in a PPKTP waveguide in a more indirect manner, by measuring, on the one hand, its marginal distributions and, on the other hand, the visibility of the polarization Hong-Ou-Mandel (HOM) interference~\cite{Rubin} in the presence of more 'soft' filtering, with the FWHM on the same order of magnitude as the TPSA width. It is worth noting that, while HOM polarization interference has been already observed for cw-pumped SPDC in PPKTP waveguides~\cite{fiorentinoVISIBILITY}, our experiment is the first observation of this type of interference in waveguides pumped by femtosecond pulses.

Consider type II quasi-phasematched SPDC in a PPKTP waveguide pumped by a femtosecond-pulse pump. The waveguide is along the x axis, the pump and idler waves are assumed to be y-polarized, while the signal wave is polarized along the z axis. The two-photon state can be written in the form~\cite{Keller},~\cite{Grice}

\begin{equation}
|\Psi\rangle=\int \hbox{d}\omega_s\hbox{d}\omega_i F(\omega_s,\omega_i)a^{\dagger}(\omega_s)a^{\dagger}(\omega_i)|\hbox{vac}\rangle,
\label{state}
\end{equation}
with the two-photon amplitude $F(\omega_s,\omega_i)$ given by the product of the spectral amplitude $E_p(\omega)$ of the pump intensity and the 'phase-matching function', determined by the properties of the nonlinear crystal,

\begin{eqnarray}
F(\omega_s,\omega_i)=E_p(\omega_s+\omega_i-\omega_p)\nonumber\\
\times\hbox{sinc}\left[\frac{L}{2}\{k_p(\omega_s+\omega_i)-k_s(\omega_s)-k_i(\omega_i)-mK\}\right].
\label{TPA}
\end{eqnarray}
Here, $\omega_{p,s,i}$ are frequencies of the pump, signal, and idler radiation, respectively; $k_{p,s,i}$ are the corresponding wavevectors, $L$ is the length of the waveguide, and the 'sinc-function' is defined as $\hbox{sinc}(x)\equiv \hbox{sin}(x)/x$. The term $mK$ is due to the $m$-th order quasi-phasematching, with the inverse grating vector being related to the poling period $\Lambda$ as $K=2\pi/\Lambda$. Because of the waveguide propagation, the transverse wavevector mismatch does not enter Eq.(\ref{TPA}), and the pump, signal, and idler spatial modes are assumed to be the eigenmodes of the waveguide. Taking into account the dispersion dependence of bulk KTP as well as the waveguide-mode correction to the refractive indices~\cite{deltaNinWG}, we calculated the TPSA for 1st-order quasi-phasematching in a waveguide with $\Lambda=8.52 \mu \hbox{m}$. As a function of signal and idler wavelengths, the two-photon amplitude is shown in Fig.1. The pump wavelength was taken to be $397.65$ nm, and the width of the pump spectrum, $4.5$ nm.

\begin{figure}
\includegraphics[width=0.45\textwidth]{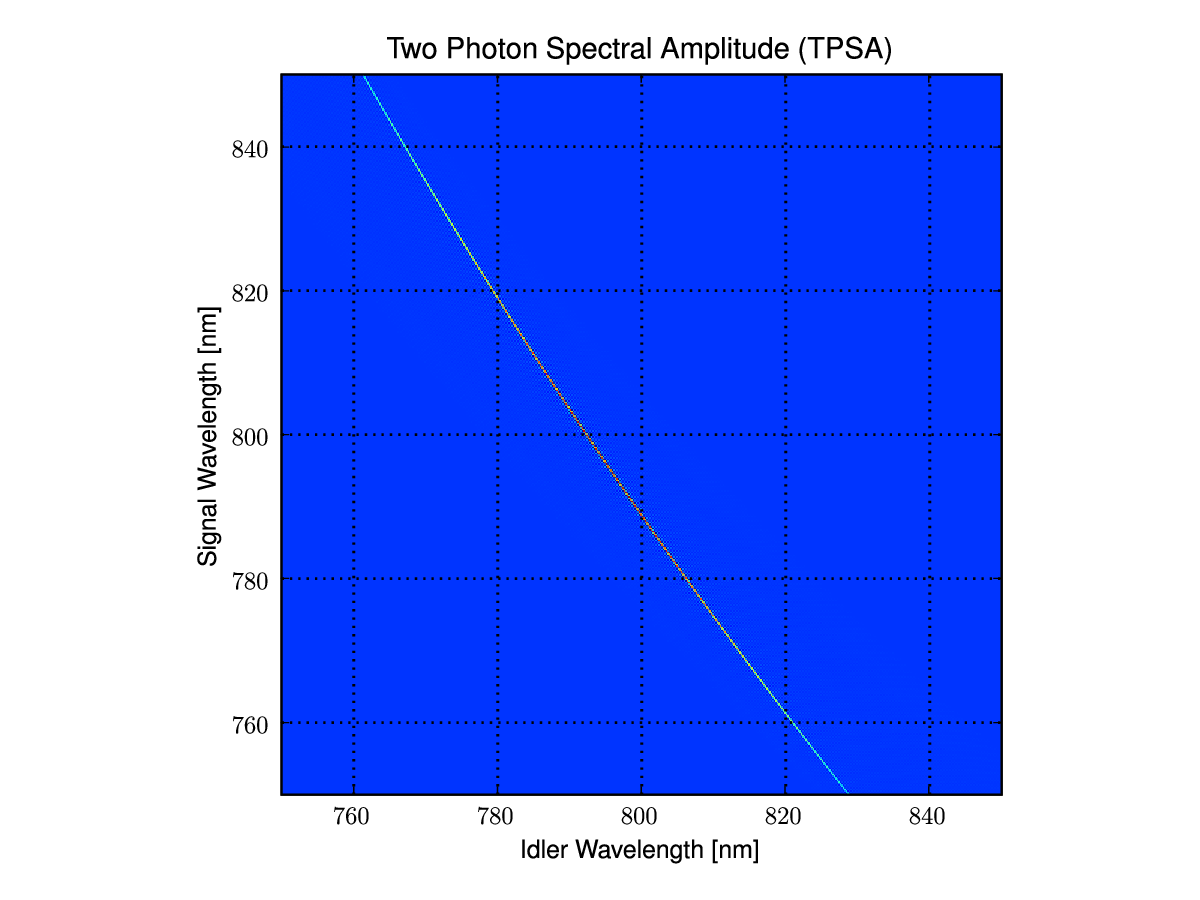}
\includegraphics[width=0.45\textwidth]{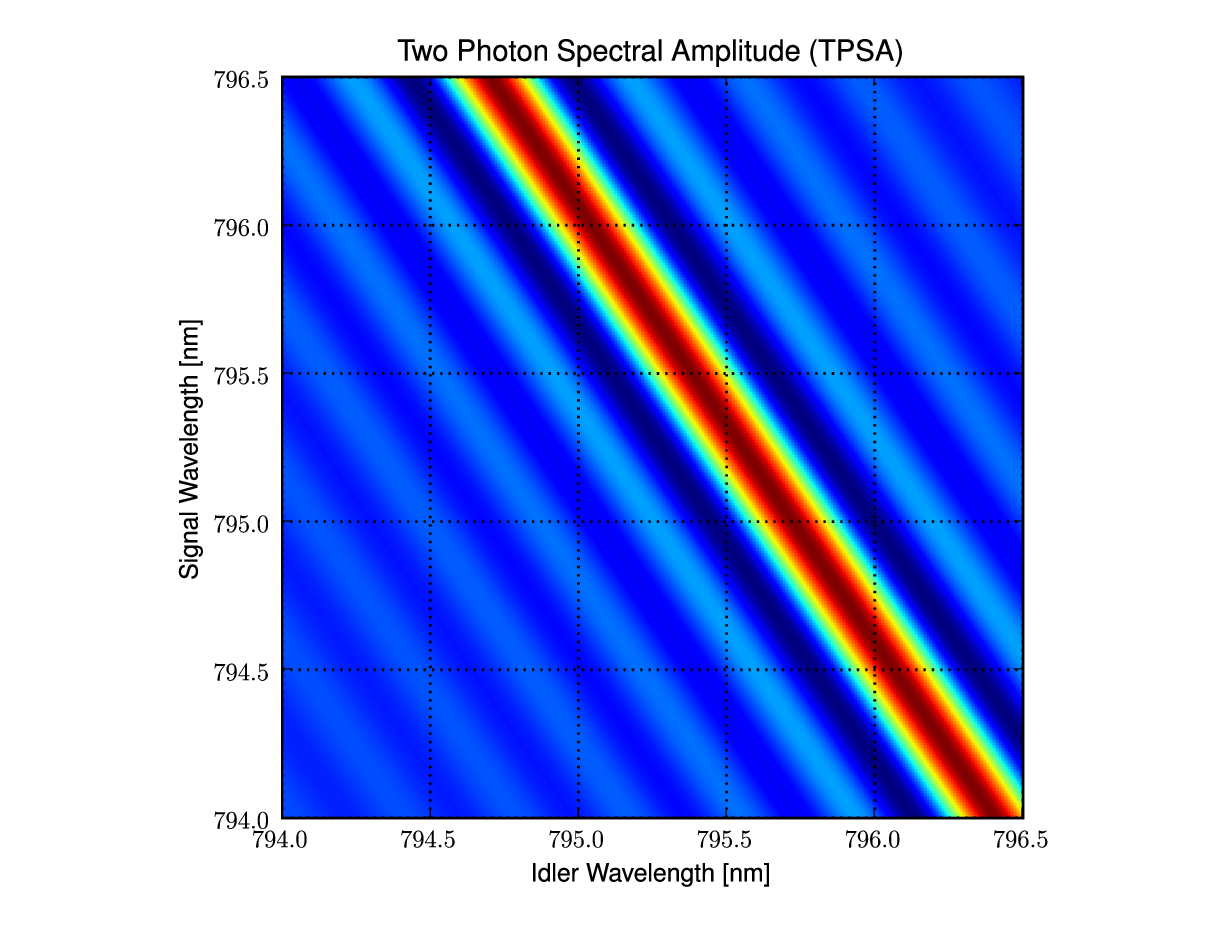}
\caption{TPSA on a large scale (a) and after a zoom into its center (b)} \end{figure}

From Fig.1a, showing the shape of the distribution, and from Fig.1b, zooming into its center, one can see that,  first, the distribution is apparently asymmetric and, second, the state is highly entangled in frequency. Indeed, while the 'marginal' widths of this distribution along signal and idler wavelengths are, respectively, $60$ and $40$ nm (which agrees well with the results of~\cite{uren}), conditional widths are much narrower: $0.26$ nm and $0.175$ nm. According to the entanglement measure $R$ suggested in Ref.\cite{Fedorov}, this results in $R=230$. To the best of our knowledge, this is the first time such a high degree of entanglement is reported for a type-II source, and it is due to the large crystal length, on the one hand, and to the group-velocity dispersion, on the other hand~\cite{Fedorov_freq}.

In order to verify these results in experiment, we made two kinds of measurements. Marginal distributions of the two-photon amplitude were measured with a spectrometer. At the same time, to estimate the width of the conditional distribution of the two-photon amplitude, we used the technique of polarization HOM interference. Because of the asymmetry of the two-photon amplitude, the interference visibility depends on the degree of entanglement: the visibility is determined by the 'relative overlap' function~\cite{Grice},

\begin{equation}
O\equiv\frac{\iint\hbox{d}\omega_s\hbox{d}\omega_i F(\omega_s,\omega_i)F^{*}(\omega_i,\omega_s)}{\iint\hbox{d}\omega_s\hbox{d}\omega_i |F(\omega_s,\omega_i)|^2},
\label{overlap}
\end{equation}
which reduces with the increase of $R$. By using filters of a sufficiently small bandwidth, inserted into both signal and idler channels, one can reduce the degree of entanglement and hence increase the 'overlap' $O$, which, in its turn, leads to an increase in the visibility.

For a 'traditional' scheme of polarization HOM interference measurement, in which a non-polarizing beamsplitter (NPBS) is followed by two polarizers~\cite{Rubin}, the visibility is equal to the overlap $O$. Here, we use a different scheme, in which the NPBS is replaced by a polarizing beamsplitter (PBS), and the coincidence counting rate is recorded versus the orientation of a half-wave plate inserted before the PBS. In this case, one can show that the visibility is related to the overlap function as
\begin{equation}
V=\frac{1+O}{3-O}.
\label{visibility}
\end{equation}
One can see that the visibility changes from $1/3$ (the classical limit) to $1$ as the overlap changes from $0$ to $1$.

Our experimental setup is shown in Fig. 2. A Ti-sapphire laser
with the central wavelength \unit{795}{\nano\meter}, the pulse duration 160~fs, and a
repetition rate of \unit{80}{\mega\hertz} was frequency doubled and used to pump a
\unit{12}{\milli\meter} long PPKTP chip. The pump was coupled into the
waveguide by using a $20\times$ achromatic microscope objective, mounted
on a 3D micrometer stage. On the surface of the PPKTP chip (from AdVR Corp.)
there were 56 waveguide channels of the size
4$\times$4~{\micro\meter\squared} with different poling periods,
separated by \unit{25-50}{\micro\meter}. The one used in our experiments had $\Lambda=8.52\mu $m. A $20\times$ achromatic objective
at the end of the waveguide chip was used to couple out the SPDC and
the transmitted pump beam. The remaining part of the pump beam was eliminated by a
high-reflectance UV-mirror, which at the same time had 99\% transmission
of the SPDC around \unit{795}{\nano\meter}. For the alignment purposes we also
monitored the spatial mode of the reflected pump beam by a CCD camera.

\begin{figure} \includegraphics[width=0.3\textwidth]{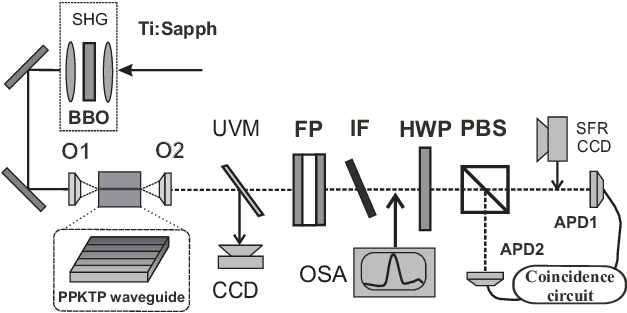}
\caption{Experimental setup. Second harmonic of Ti:Sa laser radiation is focused into a 12 mm type-II PPKTP waveguide, where SPDC
is produced; O1 and O2 are the coupling/decoupling
objectives; UVM, highly reflective UV mirror; HWP, half-wave plate; PBS,
polarizing beamsplitter; APD1, APD2, single-photon counting modules;
FP, Fabry-Perot interferometer; OSA, optical spectrum analyzer; IF,
 interference filters; SPR CCD, single-photon resolving CCD camera combined with a grating spectrometer.} \end{figure}

Observation of polarization HOM interference was performed
with a rotating half-wave plate (HWP) followed by a polarizing beam splitter
and two fiber-coupled avalanche photodiodes (SPCM from Perkin\&Elmer).
The output pulses from the detectors were electronically time-gated
by a trigger signal from the laser and then
fed into a coincidence detection setup with a resolution
window of \unit{5}{\nano\second}. By recording the coincidences versus the orientation of the HWP we characterized the visibility of the two photon HOM inferference.

In addition to our HOM experiment we measured the marginal spectral distributions by using a single photon sensitive CCD camera (Andor tech.) accompanied by a grating spectrometer with a standard resolution of 0.5nm for the signal and idler photons.
Although the maxima of two marginal distributions were slightly
shifted (about \unit{1}{\nano\meter}), we observed a region of their strong
intersection, which we further considered as a point of frequency degeneracy. The widths and the central positions of the marginal distributions agreed well with the calculation.

The main part of the experiment was dedicated to studying the effect of spectral
filtering on the visibility of HOM polarization interference. In the absence of narrowband filtering, by recording the number of coincidences
versus the orientation of the HWP, we observed a visibility of 0.33, which is the classical limit for our scheme. Spectral filtering performed only by
an IF with \unit{3}{\nano\meter} FWHM led to a visibility of $0.38$, see Fig.~3a. This is in perfect agreement with the theoretical prediction assuming the filter to have the transmission function with super-Gaussian shape
and a width of $3$ nm.
In order to perform a
strong spectral selection we installed a low-finesse ($\mathcal{F}=7$) Fabry-Perot (FP)
interferometer (LOMO, insertion losses \unit{15}{\deci B}), accompanied by
a set of narrow-band interference filters.
We stress that the same
spectral selection was performed for both signal and idler photons since the filter was
placed before the PBS. Using the base (the spatial separation
between the two mirrors) equal to \unit{300}{\micro\meter}, we achieved a nearly Lorentzian transmission spectrum of width \unit{0.15}{\nano\meter}. The transmission spectrum was probed by passing a part of the fundamental bright beam directly
through the filter and into an optical spectrum analyzer (OSA) with a resolution of \unit{0.05}{\nano\meter}(Fig.~4).
The transmission sidebands of FP were suppressed by
a combination of two interference filters with the central wavelength at
\unit{808}{\nano\meter} (at normal incidence) and FWHM of
\unit{3}{\nano\meter}. In order to center the transmission spectrum of the IFs around the SPDC degeneracy point, the filters were tilted by $10-12$~{\degree}. The
central peak corresponds to the degeneracy point, while the less pronounced
sideband of FP is due to the imperfect suppression by two IFs.

The dependence of the coincidences count rate on the  HWP orientation in the case of
filtering with the bandwidth of \unit{0.15}{\nano\meter} is shown in Fig.~3b. In order to
overcome the insertion losses introduced by the filters, we set the input mean pump
power to \unit{4}{\milli\watt}.
In this case
the contribution of the accidental coincidences was noticeable and could be estimated as
$R_{\text{acc}}={R_1R_2}/{R_{\text{rep}}}$, where $R_1$ and $R_2$ were the
counting rates of two detectors and $R_{\text{rep}}$, the repetition rate of
the pump. In the presented experiments the contribution of $R_{\text{acc}}$ was
typically at a level of {5-7\%}. The obtained visibility value, after the
subtraction of accidental coincidences, was 0.79. This value, as well as the previously mentioned results, is shown in Fig.5, together with the theoretically calculated dependence of the visibility on the filter bandwidth (solid line). One can see that the experimental points are in good agreement with the theoretical calculation, which confirms the extremely large degree of frequency entanglement achieved for waveguided type-II SPDC in PPKTP. For comparison, Fig.5b also shows the visibility dependence on the filter bandwidth calculated for the case of the two-photon amplitude conditional distribution being twice as broad (dashed line) and twice as narrow (dotted line) as given by our calculation.


\begin{figure} \includegraphics[height=4cm]{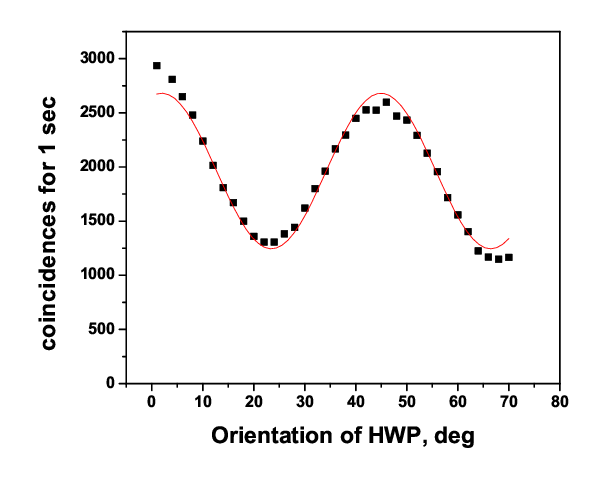}
\includegraphics[height=4cm]{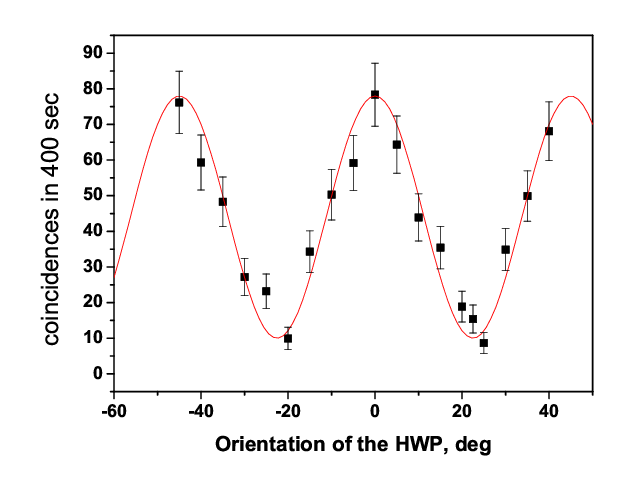}
\caption{Experimental dependencies
of the coincidence counting rate on the HWP orientation for (a)
\unit{3}{\nano\meter} filtering and (b)
\unit{0.15}{\nano\meter} filtering. Solid lines represent a fit of experimental data.} \end{figure}

\begin{figure} \includegraphics[width=0.3\textwidth]{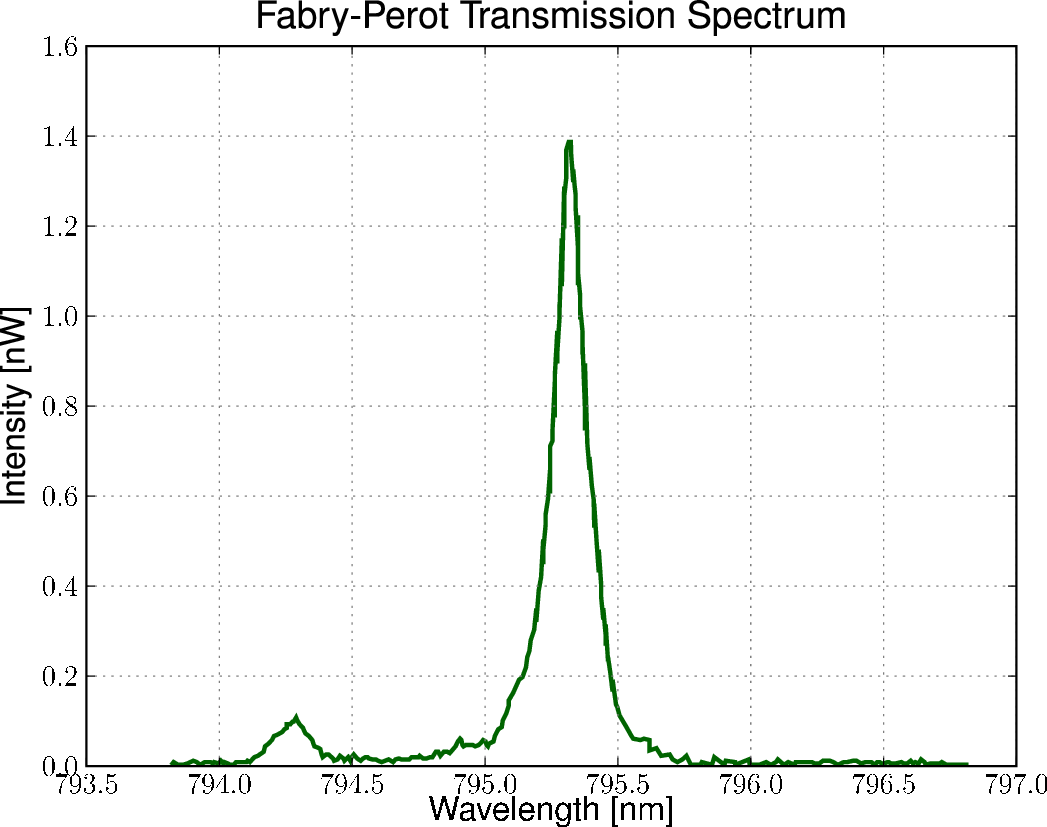} \caption{Transmission spectrum of a FP with \unit{300}{\micro\meter} base.
The central peak corresponds
to the degeneracy point, the side one is due to its
imperfect suppression by two IFs} \end{figure}

\begin{figure} \includegraphics[width=0.3\textwidth]{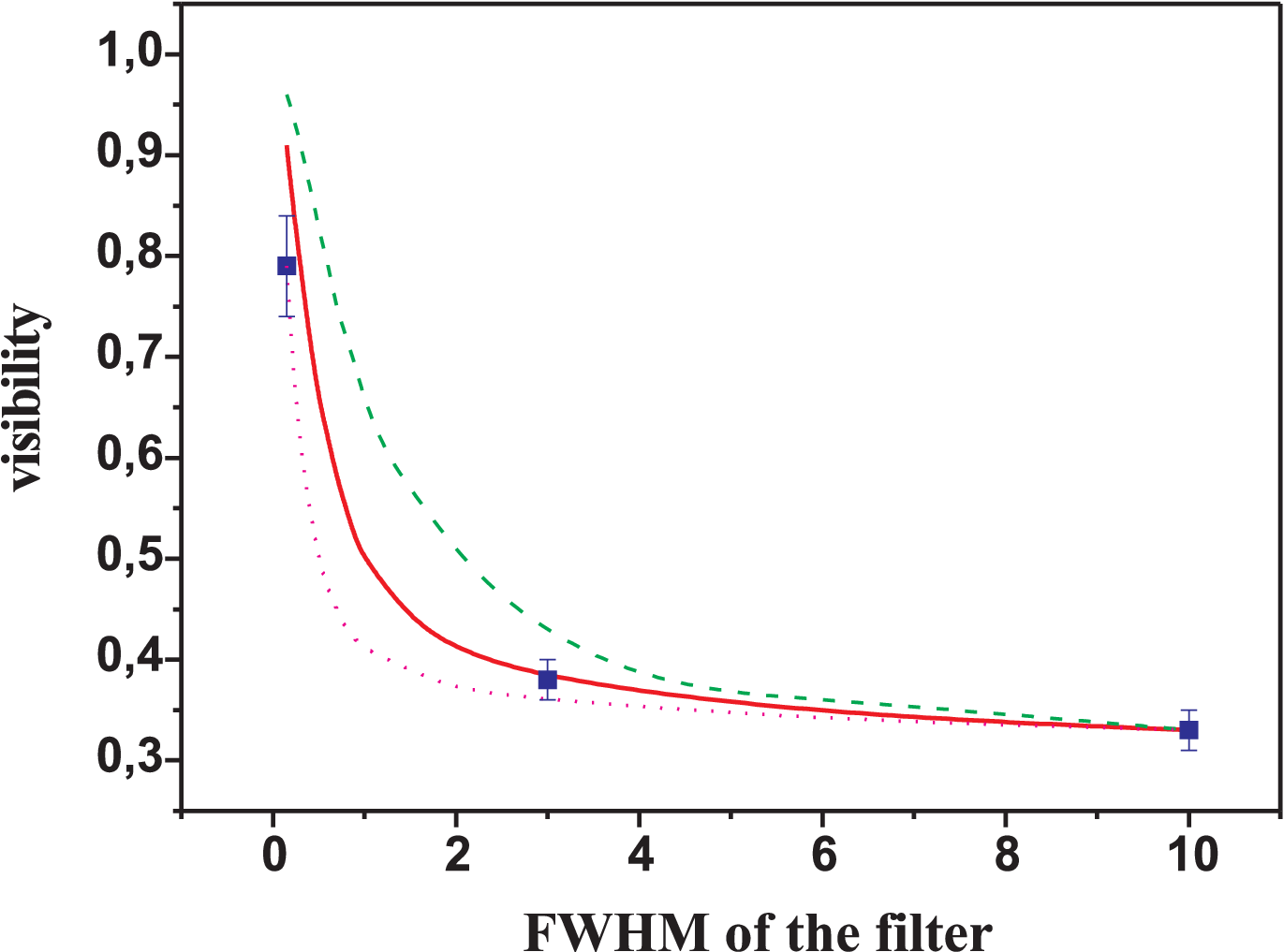} \caption{Interference visibility measured (points) and calculated (solid line), as a function of the filter bandwidth. Dashed line shows the same dependence calculated for the case of a two-photon state twice less entangled (TPSA twice as broad) and dotted line, for the case of a state twice more entangled (TPSA half as broad).} \end{figure}

In conclusion, we have studied, both theoretically and experimentally, the
distribution of the TPSA for light generated via type-II
SPDC in a PPTKP waveguide in the case of femtosecond-pulsed pump.
This distribution demonstrates a high (more than 200) degree of frequency correlation.
This means strongly multimode behavior and results in a low interference visibility for a polarization HOM-type experiment. By using narrowband filtering, with the help of a Fabry-Perot interferometer in combination with two IF filters, we have demonstrated, for the first time, high-visibility polarization interference for
ultrafast-pumped SPDC in PPKTP waveguides.

While being a drawback in some techniques (like single-mode Fock-state preparation), a high degree of frequency entanglement can be a great advantage in other applications, as it enables using high-dimensionality Hilbert space. In particular, biphoton states with high degree of frequency entanglement can form a basis for quantum key distribution protocols using continuous variables of single photons~\cite{QKD}.

We acknowledge the financial support of the Future and Emerging Technologies (FET) programme within the Seventh Framework Programme for Research of the European Commission, under the FET-Open grant agreement CORNER, number FP7-ICT-213681. L.A.K. acknowledges the support of
Alexander von Humboldt foundation, and M.V.C., the support of the DFG foundation.


\begin{references}


\bibitem{NC} M.A.~Nielsen and I.L.~Chuang, "Quantum Computation and
Quantum Information", Cambridge Univ. press (Cambridge, 2000). D.
Bouwmeester et al., "The physics of Quantum Information", Springer
Verlag (Berlin, 2000).

\bibitem{Kok} P.~Kok et al., Rev Mod. Phys. 79 (2007) 135.

\bibitem{gis} N.~Gisin et al. 2002 \emph{Rev. Mod. Phys} \textbf{74} 145.

\bibitem{alan} A.~Migdall, Phys. Today \textbf{52}, 41-46 (1999).

\bibitem{DNK} D.N.~Klyshko, \emph{Photons and Nonlinear Optics} (Gordon and
Breach, New York, 1988).

\bibitem{sergienko} Y.H.~Shih and A.V.~Sergienko, Phys. Rev. A \textbf{50},
2564 (1994).

\bibitem{Kwiat_rings} P.G.~Kwiat et al.,
Phys. Rev. Lett. \textbf{75}, 4337-4341 (1995).

\bibitem{kurt} C.~Kurtsiefer et al., Phys. Rev. A \textbf{64}, 023802 (2001).

\bibitem{Kwiat} P.G.~Kwiat et al.,
Phys. Rev. A
\textbf{60}, R773-R776 (1999).

\bibitem{kulik} Y.H.~Kim, S.P.~Kulik, and Y.H.~Shih, Phys. Rev. A \textbf{63},
060301(R) (2001).

\bibitem{gisinPPLN} S.~Tanzilli et al., Electron. Lett. \textbf{37}, 28 (2001).

\bibitem{Walmsley} K.~Banaszek et al., Opt. Lett. 26, 1367-1369  (2001);
A.~Fedrizzi et al., Opt. Exp. 15, 15377–15386 (2007).

\bibitem{fiorentino} C.E.~Kuklewicz et al.,
Phys. Rev. A \textbf{69}, 013807 (2004).

\bibitem{uren} A.B.~U'Ren et al.,
Phys. Rev. Lett. \textbf{93}, 093601 (2004);

\bibitem{deltaNinWG} M. G.~Roelofs et al.,
J. Appl. Phys. \textbf{76}, 4999 (1994).


\bibitem{Fedorov} M.V.~Fedorov et al.,
Phys. Rev. A \textbf{69}, 052117 (2004).

\bibitem{Antia} H. Sh.~Poh et al.,
Phys. Rev. A \textbf{75}, 043816 (2007).

\bibitem{Rubin} M.H.~Rubin et al.,
Phys. Rev. A \textbf{50}, 5122 (1994).

\bibitem{fiorentinoVISIBILITY} M.~Fiorentino et al. OPTICS EXPRESS
\textbf{15}, 12, 7479 (2007).

\bibitem{Keller}T.E.~Keller and M.H.~Rubin, Phys. Rev. A \textbf{56}, 1534-1541 (1997).

\bibitem{Grice}W.P.~Grice and I.A.~Walmsley, Phys. Rev. A  \textbf{56}, 1627-1634 (1997).

\bibitem{Fedorov_freq} Yu.M.~Mikhailova, P.A.~Volkov, M.V.~Fedorov, arXiv:0801.0689v1 [quant-ph] (2008).

\bibitem{QKD} I.~Ali Khan and J.~C.~Howell, Phys. Rev. A \textbf{73}, 031801 (R) (2006); L.~Zhang, Ch.~Silberhorn, and I.~A.~Walmsley, Phys. Rev. Lett. \textbf{100}, 110504 (2008).






\end{references}
\end{document}